\documentclass{article}

\newcommand{\position}{\mathbf{r}}
\newcommand{\momentum}{\mathbf{p}}
\newcommand{\angularmomentum}{m\position\wedge\dot{\position}}
\newcommand{\potential}{\mathbf{A}}

\title{Remarks on the Theory of Angular Momentum}
\author{O. Chavoya-Aceves\\Camelback HS\\Phoenix, AZ USA}
\begin{document}
\maketitle
\abstract{A rigorous application of the correspondence rules shows that the
operator of the angular momentum of a quantum particle---corresponding to the
classical magnitude $\mathbf{l}=\angularmomentum$---is given
by $\mathbf{\hat{l}}=\mathbf{r}\wedge(-i\hbar\mathbf{\nabla}
-\frac{e}{c}\mathbf{A})$ in the presence of an electromagnetic field. Thus,
despite the general opinion on the corresponding rules of quantization, the
eigenvalues of the angular momentum depend on the configuration of the
electromagnetic field. The usual rules of commutation
$[{\hat{l}}_i,{\hat{l}}_j]=i\hbar\epsilon_{ijk}{\hat{l}}_k$, that are at the
foundation of the calculus of angular momentum and of the theory of
\emph{spin}---and Bohm's example of the EPR argument---are not valid in the
presence of an electromagnetic field. The expected value of the operator
$\mathbf{\hat{l}}=-i\hbar\mathbf{r}\wedge\mathbf{\nabla}$ is not gauge
invariant, it depends on the calibration of the electrodynamic potentials.

\noindent\textbf{Pacs: 03.53.-w} Quantum Mechanics.
}

\section{Angular Momentum in the Presence of an Electromagnetic Field}
Consider a classical particle with a mass $m$ and an electric charge $e$. The
Hamilton's Function is
\begin{equation}
 \label{hamiltonian of classical particle}
 H(\position,\momentum)=\frac{\left( \momentum - \frac{e}{c}\potential
\right)^2}{2 m} + V(\position),
\end{equation}
and, in accord with classical mechanics, 
\begin{equation}
 \label{momentum of a classical particle}
 m \dot{\position} = \momentum - \frac{e}{c}\potential,
\end{equation}
indicating that
\begin{equation}
 \label{angular momentum of classical particle}
 \angularmomentum = \position \wedge \left( \momentum - \frac{e}{c}\potential
\right).
\end{equation}

Therefore, if the \emph{correspondence principle} is rigorously applied
\cite[p. 68]{MESSIAH}, the quantum operators of angular momentum, from which
the magnetic moment can be obtained, are
\begin{equation}
 \label{operator of angular momentum}
 {\hat{L}}_i = {\hat{l}}_i - \frac{e}{c}\epsilon_{ijk}x_j A_k,
\end{equation}
where
\begin{equation}
 \label{common operator of angular momentum}
 {\hat{l}}_i = - i\hbar \epsilon_{ijk}x_j\partial_k
\end{equation}
are the common operators.

To get the commutation relations we will use the identities
\begin{equation}
 \label{common commutation relations}
 [{\hat{l}}_{i},{\hat{l}}_{j}] = i\hbar\epsilon_{ijk}{\hat{l}}_{k} 
\end{equation}

and
\begin{equation}
 \label{auxiliary commutation relation}
 [{\hat{l}}_{i},\phi] = -i\hbar \epsilon_{ijk}x_j\partial_k \phi
\end{equation}

for any function $\phi$, since
\[
  -i\hbar \epsilon_{ijk}x_j\partial_k (\phi \psi)  
  + \phi i\hbar \epsilon_{ijk}x_j\partial_k  \psi = -i\hbar
\epsilon_{ijk}x_j(\partial_k \phi )\psi
\]
for any function $\psi$.

From (\ref{operator of angular momentum})
\begin{equation}
\label{expansion of the commutator of the components of angular momentum}
[{\hat{L}}_i,{\hat{L}}_j]=\left[{\hat{l}}_i - \frac{e}{c}\epsilon_{icd}x_c
A_d,{\hat{l}}_j - \frac{e}{c}\epsilon_{jcd}x_c A_d \right]
\end{equation}
\[
 = \left[{\hat{l}}_i,{\hat{l}}_j  \right]
- \frac{e}{c}\epsilon_{jcd}\left[{\hat{l}}_i ,x_c A_d \right]
- \frac{e}{c}\epsilon_{icd} \left[ x_c
A_d,{\hat{l}}_j  \right]
+\left(\frac{e}{c}\right)^2 \epsilon_{icd} \epsilon_{jcd}  \left[ x_c
A_d, x_c A_d \right]
\]
\[
 = \left[{\hat{l}}_i,{\hat{l}}_j  \right]
- \frac{e}{c}\epsilon_{jcd}\left[{\hat{l}}_i ,x_c A_d \right]
- \frac{e}{c}\epsilon_{icd} \left[ x_c
A_d,{\hat{l}}_j  \right].
\]
(The last equation follows from the fact that
\[
 \left[ x_c A_d, x_c A_d \right] \equiv 0.)
\]
Using (\ref{auxiliary commutation relation}):
\begin{equation}
 \label{auxiliary equation number one}
 \left[{\hat{l}}_i ,x_c A_d \right] = -i\hbar
\epsilon_{iab}x_a\partial_b(x_c A_d) =-i\hbar
\epsilon_{iab}(x_a \delta_{bc} A_d + x_a x_c\partial_b A_d)
\end{equation}
\begin{equation}
 \label{auxiliary equation number two}
  \left[ x_c A_d,{\hat{l}}_j  \right] = i\hbar
\epsilon_{jab}x_a\partial_b(x_cA_d)=i\hbar
\epsilon_{jab}(x_a \delta_{bc } A_d + x_ax_c\partial_b A_d)
\end{equation}

Combining (\ref{common commutation relations}), (\ref{expansion of the
commutator of the components of angular momentum}), (\ref{auxiliary equation
number one}), and (\ref{auxiliary equation number two}):
\begin{equation}
 \label{second representation of the commutation relations}
  [{\hat{L}}_i,{\hat{L}}_j]=i\hbar\epsilon_{ijk}{\hat{l}}_{k} 
+ \frac{i\hbar e}{c} \left(\epsilon_{iab}\epsilon_{jcd}
-   \epsilon_{jab}\epsilon_{icd} \right)(x_a \delta_{bc } A_d + x_ax_c\partial_b
A_d).
\end{equation}

Next:
\begin{equation}
 \label{auxiliary equation number three}
 \left(\epsilon_{iab}\epsilon_{jcd}
   -\epsilon_{jab}\epsilon_{icd} \right) \delta_{bc } =
  \epsilon_{iab}\epsilon_{jbd}
   -\epsilon_{jab}\epsilon_{ibd} =\epsilon_{jab}\epsilon_{idb}
-\epsilon_{iab}\epsilon_{jdb}    
\end{equation}
\[
 =(\delta_{ij}\delta_{ad}
-\delta_{jd}\delta_{ia})-(\delta_{ij}\delta_{ad}-\delta_{jd}\delta_{ia})
= \delta_{id}\delta_{ja}-\delta_{jd}\delta_{ia} = \epsilon_{ijk}\epsilon_{dak}
\]
\[
 -\epsilon_{ijk}\epsilon_{kad}.
\]
From (\ref{second representation of the commutation relations}) and
(\ref{auxiliary equation number three}):
\[
  [{\hat{L}}_i,{\hat{L}}_j]=i\hbar\epsilon_{ijk}\left({\hat{l}}_{k} 
-  \frac{e}{c}\epsilon_{kad} x_a A_d \right)
+ \frac{i\hbar e}{c}\left(\epsilon_{iab}\epsilon_{jcd}
-   \epsilon_{jab}\epsilon_{icd} \right)x_ax_c\partial_b
A_d.
\]
or
\begin{equation}
 \label{third representation of the commutation relations}
  [{\hat{L}}_i,{\hat{L}}_j]=i\hbar\epsilon_{ijk} {\hat{L}}_k
+ \frac{i\hbar e}{c}\left(\epsilon_{iab}\epsilon_{jcd}
-   \epsilon_{jab}\epsilon_{icd} \right)x_ax_c\partial_b
A_d.
\end{equation}

The tensor $x_ax_c$ is symmetric; therefore:
\[
 \left(\epsilon_{iab}\epsilon_{jcd}-\epsilon_{jab}\epsilon_{icd} \right)x_ax_c
= 
 \frac{\left(\epsilon_{iab}\epsilon_{jcd}  + \epsilon_{icb}\epsilon_{jad}
 -\epsilon_{jab}\epsilon_{icd} -\epsilon_{jcb}\epsilon_{iad}
\right)x_ax_c}{2}.
\]
Now the tensor $\frac{\left(\epsilon_{iab}\epsilon_{jcd}  +
\epsilon_{icb}\epsilon_{jad}
 -\epsilon_{jab}\epsilon_{icd} -\epsilon_{jcb}\epsilon_{iad}
\right)x_ax_c}{2}$ is antisymmetric, from which we conclude:
\[
 \left(\epsilon_{iab}\epsilon_{jcd}
-   \epsilon_{jab}\epsilon_{icd} \right)x_ax_c\partial_b
A_d 
\]\[
=\frac{\left(\epsilon_{iab}\epsilon_{jcd}  +
\epsilon_{icb}\epsilon_{jad}
 -\epsilon_{jab}\epsilon_{icd} -\epsilon_{jcb}\epsilon_{iad}
\right)x_ax_c(\partial_b
A_d -\partial_d A_b)}{4}
\]
\[
 =\frac{\epsilon_{bde}\left(\epsilon_{iab}\epsilon_{jcd}  +
\epsilon_{icb}\epsilon_{jad}
 -\epsilon_{jab}\epsilon_{icd} -\epsilon_{jcb}\epsilon_{iad}
\right)x_ax_c H_e}{4},
\]
where $\mathbf{H}=\mathbf{\nabla}\wedge\mathbf{A}$ is the magnetic field.
Furthermore:
 
\[
 \epsilon_{bde}\epsilon_{iab}\epsilon_{jcd} x_a x_c=
\epsilon_{bde}\epsilon_{bia}\epsilon_{jcd} x_a x_c= (\delta_{di}\delta_{ea}
-\delta_{da}\delta_{ei})\epsilon_{jcd} x_a x_c = \epsilon_{ijk} x_k x_e ;
\]
\[
 \epsilon_{bde}\epsilon_{icb}\epsilon_{jad}x_a x_c=\epsilon_{bde}\epsilon_{bic}
\epsilon_{jad}x_a x_c=(\delta_{di}\delta_{ec}
-\delta_{dc}\delta_{ei})\epsilon_{jad} x_a x_c= \epsilon_{ijk} x_k x_e ;
\]

\[
 \epsilon_{bde} \epsilon_{jab}\epsilon_{icd} x_ax_c  =
  \epsilon_{deb} \epsilon_{jab}\epsilon_{icd} x_ax_c
 (\delta_{dj}\delta_{ea} -\delta_{da}\delta_{ej})\epsilon_{icd} x_a x_c=
 \epsilon_{ikj}x_k x_e 
\]\[
= - \epsilon_{ijk}x_k x_e ;
\]
and
\[
 \epsilon_{bde}\epsilon_{jcb}\epsilon_{iad} x_a x_c =  
 \epsilon_{deb}\epsilon_{jcb}\epsilon_{iad} x_a x_c =
 (\delta_{dj}\delta_{ec}-\delta_{dc}\delta_{ej})\epsilon_{iad} x_a x_c  
 =
 \epsilon_{ikj}x_ex_c 
\]
\[
 = - \epsilon_{ijk}x_k x_e.
\]

In this way:
\[
\frac{\epsilon_{bde}\left(\epsilon_{iab}\epsilon_{jcd}  +
\epsilon_{icb}\epsilon_{jad}
 -\epsilon_{jab}\epsilon_{icd} -\epsilon_{jcb}\epsilon_{iad}
\right)x_ax_c H_e}{4}= (\mathbf{r}\cdot\mathbf{H})\epsilon_{ijk} x_k.
\]

On the grounds of (\ref{third representation of the
commutation relations}) and the last equation we argue that:
\begin{equation}
 \label{new commutation relation}
  [{\hat{L}}_i,{\hat{L}}_j]=i\hbar\epsilon_{ijk} \left({\hat{L}}_k +
\frac{e}{c} (\mathbf{r}\cdot\mathbf{H}) x_k \right).
\end{equation}
Thus we have proved that the commutation relations (\ref{common commutation
relations}), on which the calculus of angular momenta and the theory of
\emph{spin}---and Bohm's example of the EPR argument \cite[p.
614]{BOHM}---are grounded, are not valid in the presence of an electromagnetic
field. Anomalies in the magnetic moment must be observed in atoms and elementary
particles in the presence of magnetic fields of a high intensity.

\section{Ehrenfest Relations}
Let
\begin{equation}
 \label{operator of linear momentum}
 \pi_i = -i\hbar \partial_i - \frac{e}{c} A_i
\end{equation}
be the components of the kinetic (non-canonical) linear momentum. We have
\begin{equation}
 \label{first set of commutation relations for linear momentum}
 [\pi_i,\phi] = -i\hbar\partial_i \phi.
\end{equation}
(For any function $\phi$.)
\begin{equation}
 \label{second set of commutation relations for linear momentum}
 [{\hat{\pi}}_i,{\hat{\pi}}_j]=\frac{i\hbar e}{c}\left( 
 \partial_i A_j - \partial_j A_i
 \right) = \frac{i\hbar e}{c} \epsilon_{ijk} H_k
\end{equation}
\begin{equation}
 \label{hamiltonian operator in the presence of a magnetic field}
 \hat H = \frac{{\hat{{\pi}}}_i{\hat{{\pi}}}_i}{ 2 m } + e V.
\end{equation}

From the general formula
\begin{equation}
 \label{operador de la derivada temporal}
 {\hat{\dot{f}}} = \frac{i}{\hbar}[\hat H, \hat f] +\frac{\partial \hat
f}{\partial t}
\end{equation}
we get
\begin{equation}
 \label{derivative of the linear momentum}
 {\hat{\dot{{\pi}}}}_i = \frac{i}{\hbar}\frac{[{\hat{{\pi}}}_k{\hat{{\pi}}}_k
,{\hat{{\pi}}}_i ] }{2 m } - e\partial_i V - \frac{e}{c}\frac{\partial
A_i}{\partial t} = \frac{i}{\hbar}\frac{[{\hat{{\pi}}}_k{\hat{{\pi}}}_k
,{\hat{{\pi}}}_i ] }{2 m } + e E_i .
\end{equation}
Where $E_i$ are the components of the electric field, which allows us to
identify the operators of the components of the magnetic force:
\begin{equation}
 \label{first form of the magnetic force}
 {\hat M}_i = \frac{i}{\hbar}\frac{[{\hat{{\pi}}}_k{\hat{{\pi}}}_k
,{\hat{{\pi}}}_i ] }{2 m }.
\end{equation}

Now:
\[
  [{\hat{{\pi}}}_k{\hat{{\pi}}}_k,{\hat{{\pi}}}_i ] =
  {\hat{{\pi}}}_k{\hat{{\pi}}}_k{\hat{{\pi}}}_i -
{\hat{{\pi}}}_i{\hat{{\pi}}}_k{\hat{{\pi}}}_k =
  {\hat{{\pi}}}_k{\hat{{\pi}}}_k{\hat{{\pi}}}_i -
{\hat{{\pi}}}_k{\hat{{\pi}}}_i{\hat{{\pi}}}_k+{\hat{{\pi}}}_k{\hat{{\pi}}}_i{
\hat{{\pi}}}_k  -
{\hat{{\pi}}}_i{\hat{{\pi}}}_k{\hat{{\pi}}}_k
\]
\[
 =
  {\hat{{\pi}}}_k \left( {\hat{{\pi}}}_k{\hat{{\pi}}}_i -
{\hat{{\pi}}}_i{\hat{{\pi}}}_k \right) + \left({\hat{{\pi}}}_k{\hat{{\pi}}}_i  -
{\hat{{\pi}}}_i{\hat{{\pi}}}_k\right){\hat{{\pi}}}_k = \frac{i\hbar
e}{c}\lbrace {\hat{{\pi}}}_k , H_n  \rbrace
\]
where $\lbrace \hat A , \hat B \rbrace = \hat A \hat B + \hat B \hat A$. On
those grounds, eq. (\ref{first form of the magnetic force}) is rewritten as
\begin{equation}
 \label{second form of the magnetic force}
 {\hat M}_i = \frac{1}{2 m c} \epsilon_{ijk} \lbrace {\hat{{\pi}}}_j , H_k
\rbrace ,
\end{equation}
which is a formal translation of the Lorentz force to the language of operators.

From eqs. (\ref{operator of angular momentum}) and (\ref{operator of linear
momentum})
\begin{equation}
 \label{first expression for the torque}
  \frac{i}{\hbar}\left[ \hat H ,  {\hat L}_i \right] + \frac{\partial {\hat
L}_i}{\partial t} = \frac{i}{\hbar}\left[ \frac{{\hat{{\pi}}}_a{\hat{{\pi}}}_a}{
2 m } + e V   , \epsilon_{ijk} x_j {\hat \pi}_k \right]
-\frac{e}{c}\epsilon_{ijk} x_j \frac{\partial A_k}{\partial t}
\end{equation}
\[
 = \frac{i\epsilon_{ijk}}{\hbar} \frac{\left[ {\hat{{\pi}}}_a{\hat{{\pi}}}_a ,
x_j
{\hat{{\pi}}}_k \right]}{2m} + \epsilon_{ijk} e x_j E_k.
\]
We have that
\[
 \frac{i\epsilon_{ijk}}{\hbar} \frac{\left[ {\hat{{\pi}}}_a{\hat{{\pi}}}_a ,
x_j
{\hat{{\pi}}}_k \right]}{2m}  =  \frac{i\epsilon_{ijk}}{\hbar}
\frac{{\hat{{\pi}}}_a{\hat{{\pi}}}_a x_j
{\hat{{\pi}}}_k  - x_j {\hat{{\pi}}}_k
{\hat{{\pi}}}_a{\hat{{\pi}}}_a}{2m}
\]
and, from (\ref{first set of commutation relations for linear momentum}):
\begin{equation}
 \label{linear momentum position commutation relation}
 \left[{\hat \pi}_i, x_j \right] = - i \hbar \delta_{ij} \Rightarrow
  {\hat \pi}_i x_j  = x_j  {\hat \pi}_i - i \hbar \delta_{ij}  \mbox{ and }  
  x_j  {\hat \pi}_i = {\hat \pi}_i x_j + i \hbar \delta_{ij}.
\end{equation}
Therefore
\[
\frac{i\epsilon_{ijk}}{\hbar} {\hat{{\pi}}}_a{\hat{{\pi}}}_a x_j
{\hat{{\pi}}}_k  = \frac{i\epsilon_{ijk}}{\hbar} {\hat{{\pi}}}_a (   x_j  {\hat
\pi}_a - i \hbar \delta_{ja}   ) {\hat{{\pi}}}_k  =
\frac{i\epsilon_{ijk}}{\hbar} {\hat \pi}_a
x_j  {\hat\pi}_a {\hat \pi}_k  + \epsilon_{ijk}   {\hat\pi}_j {\hat
\pi}_k 
\]
\[
 = \frac{i\epsilon_{ijk}}{\hbar} (  x_j  {\hat \pi}_a - i \hbar \delta_{ja} 
 ) {\hat\pi}_a {\hat \pi}_k   + \epsilon_{ijk}   {\hat\pi}_j {\hat
\pi}_k  = \frac{i\epsilon_{ijk}}{\hbar} {\hat \pi}_a {\hat\pi}_a {\hat \pi}_k +
2 \epsilon_{ijk}   {\hat\pi}_j {\hat \pi}_k;
\]
\[
\frac{i\epsilon_{ijk}}{\hbar} {\hat{{\pi}}}_a{\hat{{\pi}}}_a x_j
{\hat{{\pi}}}_k  = \frac{i\epsilon_{ijk}}{\hbar} {\hat{{\pi}}}_a (   x_j  {\hat
\pi}_a - i \hbar \delta_{ja}   ) {\hat{{\pi}}}_k 
=\frac{i}{\hbar} \epsilon_{ijk}{\hat \pi }_a x_j {\hat \pi }_a{\hat \pi }_k+
\epsilon_{ijk}{\hat \pi }_j{\hat \pi }_k 
\]
 \[
 =\frac{i\epsilon_{ijk}}{\hbar} x_j {\hat\pi}_a{\hat\pi}_a{\hat\pi}_k +
2\epsilon_{ijk}{\hat\pi}_j{\hat\pi}_k;
\]
\[
 \frac{i\epsilon_{ijk}}{\hbar} \frac{{\hat{{\pi}}}_a{\hat{{\pi}}}_a x_j
{\hat{{\pi}}}_k  - x_j {\hat{{\pi}}}_k
{\hat{{\pi}}}_a{\hat{{\pi}}}_a}{2m} = \frac{i\epsilon_{ijk}}{\hbar}
x_j\frac{{\hat \pi }_a {\hat \pi }_a{\hat \pi }_k - {\hat{{\pi}}}_k
{\hat{{\pi}}}_a{\hat{{\pi}}}_a}{2m} + \epsilon_{ijk}[{\hat \pi}_j,{\hat \pi}_k];
\]
and
\[
 \frac{i\epsilon_{ijk}}{\hbar} \frac{\left[ {\hat{{\pi}}}_a{\hat{{\pi}}}_a ,
x_j
{\hat{{\pi}}}_k \right]}{2m}  = \epsilon_{ijk} x_j {\hat M}_k +
\epsilon_{ijk}[{\hat \pi}_j,{\hat \pi}_k].
\]
(See eq. (\ref{first form of the magnetic force}.) In a similar fashion we
establish that
\[
 \frac{i\epsilon_{ijk}}{\hbar} \frac{\left[ {\hat{{\pi}}}_a{\hat{{\pi}}}_a ,
x_j
{\hat{{\pi}}}_k \right]}{2m}  = \epsilon_{ijk}  {\hat M}_k x_j -
\epsilon_{ijk}[{\hat \pi}_j,{\hat \pi}_k].
\]
From eq. (\ref{first expression for the torque}) and the last two, we conclude:
\begin{equation}
 \label{final expression for the torque}
 {\hat{\dot L}}_i = \epsilon_{ijk}\lbrace x_j , {\hat f}_k \rbrace
\end{equation}
where
\begin{equation}
 \label{final expression for the lorentz force}
 {\hat f}_k = {\hat M_k} + e E_k,
\end{equation}
is the operator for the Lorentz Force, in complete harmony with the
\emph{correspondence principle}.

Notice that, the commutators $[\hat H, {\hat l}_i]$ cannot provide the correct
component of the torque associated to the electric field, which definitely
proves that the operators of magnetic moment are not proportional to ${\hat
l}_i$; the last operators do not correspond to the kinetic angular momentum.
Besides, the expected value of those operators is not invariant under gauge
transformations \cite[p. 100]{WEYL}.

\end{document}